\documentclass[prb,showpacs,twocolumn,amsmath,amssymb]{revtex4}
\usepackage{epsfig}
\usepackage{graphicx}
\usepackage{dcolumn}
\usepackage{bm}

\newcommand{\la}{\langle}
\newcommand{\ra}{\rangle}

\begin{document}
\title{Electron transmission between normal  and   
 heavy electron metallic phases in  Kondo lattices}
\author{M. A. N. Ara\'ujo$^{1,2}$\footnote{On leave from Departamento de F\'{\i}sica, 
Universidade de \'Evora, P-7000-671, \'Evora, Portugal }
  and A. H. Castro Neto$^{3}$}
\affiliation{$^1$ Department of Physics, Massachusetts Institute of Technology, Cambridge MA 02139, U.S.A.}
\affiliation{$^2$ CFIF, Instituto Superior 
T\'ecnico, Av. Rovisco Pais, 1049-001 Lisboa, Portugal}
\affiliation{$^3$Department of Physics, Boston University, 590 
Commonwealth Avenue, Boston, MA 02215,USA}

\begin{abstract}
The interface between a  heavy fermion metallic phase and a ``normal'' (light-fermion) metal phase  is discussed. 
The Fermi momentum mismatch between the two phases causes the carriers to scatter at the interface. 
The interface conductance is a monotonous increasing function of  conduction electron density, $n_c$, 
and is almost 60\% of that of a  clean heavy fermion metal at half-filling ($n_c=1$) and can be measured
experimentally. Interface experiments can be used as probe of the nature of the homogenous heavy-fermion 
state and provide important information on the effects of inhomogeneities in heavy-fermion alloys.
\end{abstract}

\maketitle

\section{Introduction}

In Kondo lattice metals a Fermi sea of carriers interacts antiferromagnetically with a lattice of localized $f$ electrons, 
in rare earth atoms, which behave as
 localized magnetic moments. The interplay between 
magnetic interactions (such as RKKY) and the Kondo effect leads to two very different types of ground states:
a magnetically ordered metal and a paramagnetic disordered  heavy metal phase. 
On one hand, in the disordered phase the $f$ moments effectively 
decouple from the conduction band which has a small Fermi sea (FS) made out of $n_c$ electrons per unit cell. 
In the heavy-fermion liquid (HFL) phase, on the other hand, the Kondo effect drives the formation of singlets between
conduction and $f$ electrons and, therefore, leads to magnetic screening and a large FS with $n_c+1$
electrons \cite{newns}. Hence, on a quantum phase transition ($T=0$) between a heavy-fermion and a
light-fermion state the metal undergoes an abrupt change in the volume of the FS. This volume change
can be observed in measurements of the Hall coefficient which is a direct measure of the density of
carriers in the system \cite{pashen}. Nevertheless, these measurements are often complicated by
extrinsic effects such as the presence of disorder and finite temperature broadening.

In the presence of disorder, either due to alloying or extrinsic impurities, the situation becomes more complex and, due to local variations of chemical pressure,
 the system may break up into domains of the two phases leading to the formation of {\it internal interfaces}  and to non-Fermi liquid behavior generated by quantum
 Griffiths singularities \cite{grif}. A similar situation occurs close to
a heavy-metal to Anderson insulator transition \cite{vlad}. It has been argued recently \cite{neto,millis} that these interfaces control the amount of dissipation and
 the crossover energy scales that regulate the physical properties of disordered alloys. Therefore,
our studies also have  implications on the study of non-Fermi liquid phases \cite{greg}.

\begin{figure}[htb]
\centerline{\includegraphics[width=6.5cm]{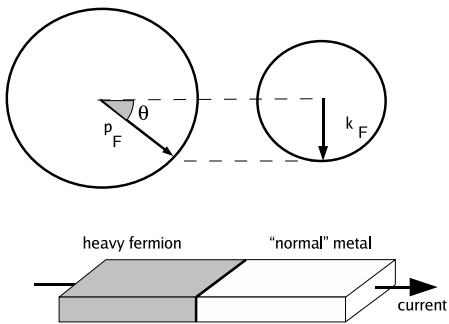}}
\caption{\label{junction}  Top: Fermi surfaces of a  heavy Fermi liquid (HFL) metal (left)
and a ``normal'' light fermion metal (right). Bottom: an experimental setup where the HFL 
and ``normal'' metal  coexist in the same sample and an electrical current flows through.}
\end{figure}

Now imagine that 
one constructs  an interface (or junction) between these two distinct
phases. 
The difference in FS volume (or Fermi momentum, $k_F$, mismatch) leads to
 an impedance at the interface and hence to a contribution to the conductance
in the system (see Fig.~\ref{junction}). Similar experiments have been
realized recently in point contact spectroscopy of normal metals and heavy
fermion superconductors such as CeCoIn$_5$, with unusual behavior of the
point contact conductance as a function of the applied voltage \cite{greene}.
We analyze below a inhomogeneous system in which a heavy fermion phase and a ``normal'' (light-fermion) metallic phase in the same
material are in direct contact with each other. Because the Kondo singlet formation lowers the energy of the conduction electrons, some electrons migrate from the normal
phase into the HFL phase, creating an electric dipole barrier between the two phases. At the interface the Fermi surface volume changes abruptly, so the interface  itself
behaves as a scatterer.    We establish the matching conditions on the electronic states  and calculate the 
electrical conductance across the interface in a simple Kondo lattice  model. 

We also provide suggestions for experiments where  two phases are made to coexist in a
single sample, as in  Fig.~\ref{junction}. 
The total resistance is given by the sum of the bulk resistances of the phases {\it plus} the interface contribution.
If the specimen is a heavy fermion material belonging  to the class of quantum critical points (QCPs) 
where the Kondo temperature vanishes, the interface scattering should be observable. If the material belongs 
to the ``SDW scenario'' of QCP, the interface contribution should be absent as there is no FS volume mismatch at the QCP. 
Therefore, such an experiment could distinguish the two types of QCPs. 

The paper is organized as follows. The model Hamiltonian is introduced in Section \ref{modelHam}; Section \ref{III} is devoted to the discussion
of the  electronic properties of the interface; experimental suggestions are given in section \ref{discuss}.

\section{Model Hamiltonian}\label{modelHam}

We consider a Kondo lattice model for the metal, with a single band $\epsilon(\bm k)$ of  conduction ($\hat c_{\bm k, \sigma}$)
 electrons with Bloch momentum $\bm k$ and spin $\sigma=\pm 1/2$, interacting antiferromagnetically with the localized $f$ electrons in a 
lattice with  $N_s$ sites:
\begin{eqnarray}
\hat H&=& \sum_{{\bm k},\sigma} \left(\epsilon(\bm k) -\mu\right)\hat c_{{\bm k},\sigma}^{\dagger}\hat c_{{\bm k},\sigma}
+ \sum_{{\bm k},\sigma} \left(\epsilon_f -\mu\right)\hat f_{{\bm k},\sigma}^{\dagger}\hat f_{{\bm k},\sigma}\nonumber \\
&-& J \sum_i \hat {\bm s}_i \cdot  \hat {\bm S}_i\,,
\label{model}
\end{eqnarray}
with $J<0$, and the spin operators at site $i$ are given by
$\hat {\bm s}_i = \hat c_{i,\alpha}^\dagger \bm \sigma  \hat c_{i,\beta}$ and
$\hat {\bm S}_i = \hat f_{i,\alpha}^\dagger \bm \sigma  \hat f_{i,\beta}$ ($\bm \sigma$ are Pauli matrices). 
There is exactly one $f$ electron at each site. In order to describe the spin singlet formation between 
$c$ and $f$ electrons at every site in the 
HFL  phase,  we recast  the local spin exchange interaction as 
$\hat {\bm s}_i \cdot  \hat {\bm S}_i =  - \left(\hat \lambda_{i,\uparrow}
+ \lambda_{i,\downarrow} \right)^2 - \hat n_{c,i} - \hat n_{f,i} - \hat n_{c,i}\hat n_{f,i}$,
where $\hat n_{c(f),i}$ is the number operator  for $c$ ($f$) electron at site $i$ and
 $\hat \lambda_{i\sigma} = \frac{1}{2}\left(\hat c_{i,\sigma}^\dagger \hat f_{i,\sigma} + h.c. \right)$.
We decouple the $\hat \lambda$ operators by introducing an auxiliary field, $\lambda$, so  the effective 
mean field theory becomes \cite{newns}:
\begin{eqnarray}
\hat H&=& \sum_{{\bm k},\sigma} (\epsilon(\bm k) -\mu)\hat c_{{\bm k},\sigma}^{\dagger}\hat c_{{\bm k},\sigma}
+ \sum_{{\bm k},\sigma} (\epsilon_f -\mu)\hat f_{{\bm k},\sigma}^{\dagger}\hat f_{{\bm k},\sigma}\nonumber \\
&-& 2\lambda J \sum_{i,\sigma} \left(\hat c_{i,\sigma}^\dagger \hat f_{i,\sigma} + h.c. \right) - 4\lambda^2 J N_s\,.
\label{model2}
\end{eqnarray}
In the HFL phase ($\lambda \neq 0$) quasi-particles form, with momentum $\bm k$, having  $c$ and $f$ components.
We shall assume that only the lowest band $E(\bm k)$ of heavy quasi-particles is occupied.
 In the  ``normal'' phase  ($\lambda =0$) quasi-particles are just $c$ electrons. 
 Single particles are  described by two-component wave-functions $\psi(\bm r)=(u(\bm r), v(\bm r))$ 
 related to the $c$ and $f$ components of the electronic states, respectively, and  obeying the normalization condition
$|u(\bm r)|^2+ |v(\bm r)|^2 =1$.  In homogeneous heavy fermion metals  we have  
$(u(\bm r),v(\bm r))  = (u_{\bm k}, v_{\bm k})  e^{i\bm k\cdot \bm r}$, with
\begin{eqnarray}
u_{\bm k}^2&=& \frac{(2\lambda J)^2}{(2\lambda J)^2 + \left(E-\epsilon(\bm k)\right)^2}\,,\\
v_{\bm k}^2&=& \frac{\left(E-\epsilon(\bm k)\right)^2}{(2\lambda J)^2 + \left(E-\epsilon(\bm k)\right)^2}\,,
\end{eqnarray} 
and the energy  of this state is
\begin{eqnarray}
E(\bm k)= \frac{1}{2} \left[
\epsilon_f + \epsilon(\bm k) \pm \sqrt{\left(\epsilon_f - \epsilon(\bm k)\right)^2  + (4\lambda J)^2}\right] \,.
\label{Eband}
\end{eqnarray} 
In a normal homogeneous metal $u_{\bm k}=1$ for a $c$ 
electron  and $v_{\bm k}=1$ for a local $f$ electron. 
The self-consistent equation for $\lambda$ is:
$
\lambda =
(4N_s)^{-1}
  \sum_{i,\sigma} \la \hat c_{i,\sigma}^\dagger \hat f_{i,\sigma} + h.c. \ra\,.
$
In  mean field theory, the local chemical potential $\epsilon_f$ for the $f$ electrons must be chosen so as to implement
single occupancy on average, $\la \hat n_{f,i}\ra =1$,  at every site $i$. Therefore, in a inhomogeneous  system $\epsilon_f$ may be
required to vary 
from site to site.
In the HFL phase $\lambda \neq 0$ and  both $c$ and $f$ electrons contribute to the FS volume. Therefore,
the volume of the FS must either include $n_c$ electrons in the normal metal or $1+n_c$ electrons in the heavy fermion metal.

We now consider a system where  both HFL and normal phases exist simultaneously and are separated by a thin interface. 
 We take half of the metal ($x<0$) to be in the  HFL phase while the other
$x>0$ half is in the normal  state ($\lambda = 0$). The interface is the $yz$ plane. We shall 
denote  by $\bm p$ the Bloch wave-vector  in the heavy metal and by $\bm k$ the wave-vector in the normal metal.
  
\section{Interface properties}\label{III}

\subsection{Absence of a proximity effect}

If we consider that $J=0$ on the normal $x>0$ half of the metal, then the amplitude for singlet 
formation, $\la \hat c_{i,\sigma}^\dagger f_{i,\sigma}\ra$, drops abruptly from its finite value in the heavy fermion ($x<0$)
metal, to zero in the normal metal \cite{next}. 
The local singlet correlation $\lambda$ vanishes abruptly
as one goes from the heavy to the normal metal because of the localized nature of the $f$ electrons. 
In this sense, we may say that there is no proximity effect for singlet formation.
In a normal-superconductor 
interface there is a proximity effect for the pairing, $\langle c_{i,\uparrow} c_{i,\downarrow} \rangle$, because the electrons 
establish the Cooper pairing in the superconductor and they both propagate into the normal side transporting the correlation with them.
This does not happen in the case discussed here since the $f$ electrons are localized.  The coupling $J=0$ on the normal side 
effectively removes the normal metal $f$ local moments from the problem, so they cannot establish singlet correlations with 
conduction electrons. Nevertheless, there is a singlet correlation between a conduction electron on the normal side (close to 
the interface) and a $f$ moment on the heavy metal side.
 The same happens if, instead of taking $J=0$, a
 sufficiently strong magnetic field  is applied to the $x>0$ portion of the metal which 
polarizes  the $f$ moments, thereby  preventing the formation of spin singlets. Alternatively,  we may consider  a temperature gradient, 
such that the left side of the sample is below the Kondo temperature, $T_K$, and the right hand side is above $T_K$.
  Then  $\la \hat c_{i,\sigma}^\dagger f_{i,\sigma}\ra$ decreases continuously to zero as the temperature crosses  $T_K$.
 The coherence factors of the quasi-particles $u$ and $v$ vary continuously in space but the FS volume of carriers changes 
abruptly in a very thin region where $\lambda \rightarrow 0$. The absence of a proximity effect is rather important in the case 
of inhomogeneous alloys because it shows that the heavy-fermion component of the system cannot penetrate the light component. The 
situation here is similar to the one in optics where light coming to an interface between two materials with very different index of
 refraction is completely reflected at the interface. 

\subsection{Formation of a dipole barrier} 

The spin singlet formation between localized $f$ and  conduction electrons lowers 
the effective local site energy of the $c$ electron. This causes the chemical potential $\mu$ of a heavy fermion system to be lower than that of a normal metal
 with the same conduction electron concentration $n_c$. It implies that some conduction electrons initially flow through the interface, leaving a positive excess 
charge in the normal metal side  and a negative excess charge on the heavy fermion side. Therefore, a dipole barrier forms close to the interface. The electrostatic
 potential created by the barrier (similar to that of a capacitor) increases the local site energy of the $c$ electrons on the heavy fermion subsystem by an amount
 $\epsilon_c$ with respect to the local $c$ site energies in the normal subsystem (the situation here resembles the contact potential of a dipole layer in p-n semiconductor 
junctions). The chemical potential is then constant throughout the sample but the $c$ site energies increase by the amount  $\epsilon_c$ across the barrier from the normal
 to the heavy fermion side.  Because these are  metallic systems, the dipole barrier should be efficiently screened by the conduction electrons over a length of the order
 of few atomic spacings \cite{next}.
\begin{figure}[htb]
\centerline{\includegraphics[width=9cm]{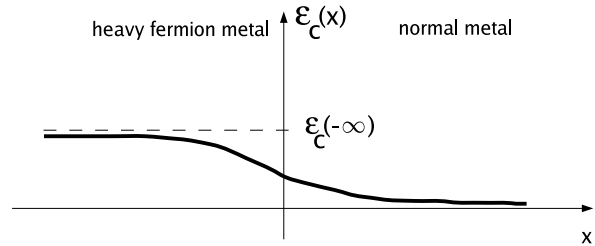}}
\caption{\label{dipole}  
Spatial variation of the  $c$ site energies due to the formation of a dipole barrier at the interface. 
}
\end{figure}

The local site  site energy of the $c$ electrons varies in space close to the interface,  $\epsilon_c(x)=eV(x)$ where $V(x)$ is the electrostatic potential and 
$e$ is the electron charge. 
It is assumed that  $V(x)\rightarrow 0$ as $x\rightarrow \infty$ deep inside the normal metal (see Figure \ref{dipole}). 
The Poisson equation for the local site energy
reads
\begin{equation}
\frac{d^2\epsilon_c(x)}{dx^2}= -\frac{e^2}{\varepsilon_0} \left[ n_c(x) -   \bar{n}_c\right]\,, \label{Poisson}
\end{equation}
where $\varepsilon_0$ is the vacuum electrical permitivity, 
$n_c(x)$ denotes the space dependent $c$ electron concentration and the value of 
$ \bar{n}_c =n_c(\pm \infty)$ is determined by charge neutrality away from the interface. 
An analytical description of the dipole barrier
can be made using a Thomas-Fermi approximation where the Fermi momenta are assumed to vary in space. 
On the heavy fermion side ($x<0$), using a parabolic dispersion, one can write:
\begin{eqnarray}
&\mu =& \frac{1}{2}\left[ \frac{\hbar^2 p_F^2(x)}{2m}+ \epsilon_c(x) + \epsilon_f \right.  \nonumber\\
&&\pm \left. \sqrt{\left(  \frac{\hbar^2 k_F^2(x)}{2m}+ \epsilon_c(x)-\epsilon_f\right)^2  + (4\lambda J)^2}\
 \right]\, ,\label{H1} \\
&n_c(x)&=\frac{2}{(2\pi)^3}\int_0^{p_F(x)}d^3p\ u_p^2(x)\,, \label{H2}
\end{eqnarray}
while on the normal metal side we write:
\begin{eqnarray}
\mu &=&  \frac{\hbar^2 k_F^2(x)}{2m}+ \epsilon_c(x)\,, \label{N1}\\
n_c(x)&=&\frac{2}{(2\pi)^3} \frac{4}{3}\pi k_F^3(x)\,. \label{N2}
\end{eqnarray}
Eliminating $ k_F(x)$ from  (\ref{N1})- (\ref{N2}) and solving for $n_c(x)$, we can then insert the result in (\ref{Poisson}) obtaining:
\begin{equation}
\frac{d^2\epsilon_c(x)}{dx^2}= -\frac{e^2}{\varepsilon_0} \left[ \left(  \frac{2m}{\hbar^2} \right)^\frac{2}{3}\frac{\left( \mu-\epsilon_c(x)\right)^\frac{3}{2}}{3\pi^2}
 -   \bar{n}_c\right]\,, 
\end{equation}
for $x>0$.
But
$$
 \bar{n}_c=  \left( \frac{2m}{\hbar^2} \right)^\frac{3}{2}\frac{\mu^\frac{3}{2}}{3\pi^2}\,,
$$
therefore,
\begin{eqnarray}
\frac{d^2\epsilon_c(x)}{dx^2} &=& -\frac{e^2}{\varepsilon_0} \frac{1}{3\pi^2}
\left(\frac{2m}{\hbar^2} \right)^{2/3}
\frac{\left[ \left(\mu-\epsilon_c(x)\right)^\frac{3}{2}-\mu^\frac{3}{2}\right]}{3\pi^2}\nonumber\\
&\approx& 
\frac{e^2}{\varepsilon_0} \left(\frac{2m}{\hbar^2} \right)^\frac{2}{3}  \frac{\mu^\frac{1}{2}}{2\pi^2}\ \epsilon_c(x)
\end{eqnarray}
implying that $\epsilon_c(x) = \epsilon_c(0) e^{-x/\xi_n}$ with the screening length inside the normal metal given by:
\begin{equation}
\xi_n^{-2}= \frac{e^2 m k_F}{\varepsilon_0 \hbar^2 \pi^2}\,.
\end{equation}

Considering the heavy fermion side ($x<0$), we may directly write the variation in $c$-electron density  already
linearized in the energy shift of the site energies: 
\begin{equation}
n_c(x)-\bar{n}_c = - N_{F(h)} \left[   \epsilon_c(x) -  \epsilon_c(-\infty)\    \right]\,,
\end{equation}
where $ N_{F(h)}$ denotes the Fermi level  density of states of the heavy fermion system (which can be taken at $x=-\infty$) and is given by
\begin{equation}
 N_{F(h)}= \frac{2}{(2\pi)^3}\int d^3p\ \delta\left(\mu - E(\bm p)\right) = \frac{m p_F}{\pi^2 \hbar^2 u^2(p_F)}\,.
\end{equation}
Then the linearized version of the Poisson  equation (\ref{Poisson}) becomes:
\begin{equation}
\frac{d^2\epsilon_c(x)}{dx^2}= \frac{e^2}{\varepsilon_0}\frac{m p_F}{\pi^2 \hbar^2 u^2(p_F)} \left[   \epsilon_c(x) -  \epsilon_c(-\infty)\    \right]\,,
\end{equation}
which gives
\begin{equation}
 \epsilon_c(x) =  \epsilon_c(-\infty) + \left[   \epsilon_c(0) -  \epsilon_c(-\infty)\    \right] e^{x/\xi_h}\,,
\end{equation}
with the screening length on the heavy metal side given by
\begin{equation}
\xi_h^{-2}= \frac{e^2 m p_F}{\varepsilon_0 \hbar^2 \pi^2u^2(p_F)}\,.
\end{equation}
The value of $p_F$ can be simply obtained from the condition of charge neutrality far from the interface,  
\begin{equation}
 \frac{2}{(2\pi)^3}\int^{p_F} d^3p\ =1 + \bar{n}_c \,.
\end{equation}

\subsection{Transmission through the interface}

The two-component wave-function $\psi(\bm r)$ varies across the interface in such a way that it describes a heavy quasi-particle
 for $x<0$ and a $c$ or localized $f$  electron for $x>0$. 
The $f$ electrons are localized and therefore carry no charge current. 
It is only the $c$ electron that transports charge (across the interface). 
This can also be readily seen from the  microscopic conservation of the probability, $\partial_t |\psi(\bm r)|^2 = -\nabla\cdot \bm j$,
 with $|\psi(\bm r)|^2 = |u(\bm r)|^2 + |v(\bm r)|^2 $. For a parabolic dispersion, $\epsilon(\nabla) \propto \nabla^2$, one obtains
the current for a $\bm k$ state as $\bm j=i \left( u \nabla u^* - u^*\nabla u\right)/\hbar$, implying that only $u(\bm r)$ carries the 
current.

The matching conditions to be imposed on the wave-function are
 the continuity of $u(\bm r)$ and the current conservation, which is only related to  $u(\bm r)$. In a model with parabolic electronic
 dispersion, $\epsilon(\bm k) \propto k^2$, this would imply the continuity of the gradient of  $u(\bm r)$. The function  $v(\bm r)$ 
itself obeys no matching condition: $v(\bm r)=1$ (or $0$) for a $f$ (or $c$) electron on the normal $(\lambda=0)$ side and,
 in the HFL side, where $\lambda \neq 0$, $v(\bm r)$ is determined by the diagonalization of the Hamiltonian (\ref{model2})
 and the matching condition for  $u(\bm r)$.

 In a model where $\lambda$ changes abruptly from a finite value to zero at the interface, the local chemical potential $\epsilon_f$ and 
the $c$
 electron site energy $\epsilon_c$ vary in space in a small region close to  the interface. In the following we neglect 
this spatial variation and assume that $\epsilon_c>0$ is constant for $x<0$ and  $\epsilon_c=0$ for $x>0$, and that   $\epsilon_f$ also
 changes abruptly
 from its constant bulk value in the heavy fermion $x<0$ to $\epsilon_f=\mu$ for $x>0$.  In order to calculate the conductance through 
the interface,
 we consider a model parabolic dispersion for the $c$ electrons. The Bloch wave-vector has both parallel ($\bm k_{||}$) and perpendicular
 ($k_x$) components 
to the interface. Introducing the position vector $\bm r = (y,z)$,  the matching conditions for the wave-function describing an incident 
quasi-particle from the 
left imply $\bm p_{||}= \bm k_{||}$:
\begin{eqnarray}
\psi_L(x<0) &=&  e^{i\cdot \bm p_{||}\cdot \bm r}\left[
 \left(\begin{array}{c}
 u \\ v \end{array}\right) e^{ip_xx} + r_{k,p} \left(\begin{array}{c} u \\ v  \end{array}\right) e^{-ip_xx}
\right]\,,   \nonumber\\
\psi_L(x>0) &=& t_{k,p}  \left(\begin{array}{c} 1 \\ 0 \end{array} \right) e^{ik_xx}  e^{i\cdot \bm p_{||}\cdot \bm r}\,,
\end{eqnarray}
with the transmission and reflection amplitudes given by:
\begin{eqnarray}
 t_{k,p} = u\frac{2p_x}{p_x+k_x}\,, \qquad
r_{k,p} = \frac{p_x-k_x}{p_x+k_x}\,,
\end{eqnarray}
respectively.  The momenta satisfy the energy conservation condition
 $E\left(\bm p_{||}, p_x\right)= \epsilon\left(\bm k_{||}=\bm p_{||},k_x\right)$. Because the heavy fermion metal has the larger
 FS only incident electrons from the left making an angle $\theta < \arcsin(k_F/p_F)$ with the $x$ axis are transmitted (see Fig.~\ref{junction}).
 A wave-function describing an incident conduction electron from the right is given by:
\begin{eqnarray}
\psi_R(x<0) &=&
t_{-p,-k} e^{i\cdot \bm k_{||}\cdot \bm r}\left(\begin{array}{c} u \\ v \end{array}\right) e^{-ip_xx}\,,\\ 
\psi_R(x>0) &=&  e^{i\cdot \bm k_{||}\cdot \bm r}\left[
\left(\begin{array}{c} 1 \\ 0 \end{array} \right) e^{-ik_xx} +
 r_{-p,-k} \left(\begin{array}{c} 1 \\ 0 \end{array}\right) e^{ik_xx}\right]\,,\nonumber
\end{eqnarray}
with the transmission and reflection amplitudes given by:
\begin{eqnarray}
 t_{-p, -k} = \frac 1 u\frac{2k_x}{p_x+k_x}\,, \qquad
r_{-p, -k} = \frac{k_x-p_x}{p_x+k_x}\,,
\end{eqnarray}
respectively.

\subsection{Conductance through the interface}
 
The transmission and reflection amplitudes are determined by the mismatch of the Fermi momenta of the two subsystems.
Applying a voltage $V$ across the interface, the charge current flowing from right to left is \cite{landauer}:
\begin{equation}
I= 2e^2V{\cal S}\int\frac{d^3k}{(2\pi)^3} \delta (\mu-\epsilon(\bm k))\  \frac{\partial E}{\hbar\partial p_x}\  |t_{-p, -k}|^2
\label{I}
\end{equation}
where ${\cal S}$ denotes the area of the interface. 
It can be shown from (\ref{Eband}) 
that the  heavy particle velocity $\partial E(\bm p)/\partial p_x=u^2 \partial \epsilon(\bm p)/\partial p_x$.
Equation (\ref{I}) can be written as:
\begin{eqnarray}
I&=&2e^2V{\cal S} 
\int_0^{\pi/2}d\theta \sin\theta\int_0^{2\pi}d\phi \int_0^\infty \frac{k^2\ dk}{(2\pi)^3}\times\nonumber\\
&&\delta\left(\hbar v_F \left(k-k_F\right)\right)\  \frac{u^2\hbar p_x}{m} \left(\frac{2k_x}{u(k_x + p_x)}\right)^2
\label{Integral}
\end{eqnarray}
 The integration is performed in the half sphere $\theta <\pi/2$ of
incident momenta from the normal metal. On the Fermi Surface we write $k_x=k_F\cos\theta$ and $p_x=p_F\cos\theta$.
 The momenta on both sides of the interface are related by
$p_F^2=\bm p_{||}^2 + p_x^2$ and $k_F^2=\bm k_{||}^2 + k_x^2$ with $p_{||}=k_{||}=k_F\sin\theta$. 
 Therefore, we can rewrite (\ref{Integral}) as
\begin{equation}
I=\frac{4e^2V{\cal S}k_F^3}{\pi h}
\int_0^{\pi/2}\frac{\sqrt{p_F^2 - k_F^2\sin^2\theta}\ \cos^2\theta\sin\theta\ d\theta}
{\left[\sqrt{p_F^2 - k_F^2\sin^2\theta} + k_F\cos\theta \right]^2}\, .
\label{Idetail}
\end{equation}

\begin{figure}[ht]
\begin{center}
\epsfxsize=7cm
\epsfbox{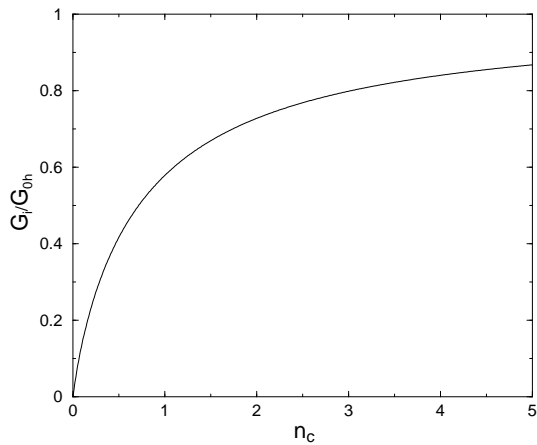}
\end{center}
\caption{Interface conductance (normalized to that of a clean heavy fermion metal)
plotted versus conduction electron density.}
\label{figura}
\end{figure}

The conductance of a clean heavy fermion metal is given by
\begin{eqnarray}
G_{0h}={\cal S}\frac{e^2}{h} \frac{p_F^2}{2\pi}  \, .
\label{gh0}
\end{eqnarray}
The conductance of the interface,  $G_i(=I/V)$, obtained from equation (\ref{Idetail}) after  changing 
the variable of integration to $x=\sin\theta$ is:
\begin{equation}
\frac{G_i}{G_{0h}}= 
8 \kappa^3 \int_0^1dx \frac{x\sqrt{1-x^2}\ \sqrt{1-\kappa^2x^2}}{\left[\sqrt{1-x^2}+\sqrt{1-\kappa^2x^2}\ \right]^2}  \, ,
\end{equation}
where $\kappa^3 = (k_F/p_F)^3= n_c/(1+n_c)$.
Figure \ref{figura} shows a plot of the interface conductance versus $n_c$. 
One can clearly see that the larger the value of $n_c$ the larger is the effect and even for dense systems
with $n_c=1$ electron per unit cell (half-filling) the value of the interface conductance is of the order of 60 \%
of that of a clean heavy fermion metal. 

\section{Experimental realization}\label{discuss}

In the geometry suggested in Fig.~\ref{junction} with the heavy-fermion, the normal fermion, and the
 interface in series, one would measure the total conductance of the junction, that is,  $G^{-1} = G_h^{-1} +G_n^{-1} + G_i^{-1}$, where
$G_n$ is the conductance of the normal Fermi liquid. Hence, in order to measure the interface conductance
one would have to measure the heavy and normal conductance separately, before measuring the conductance of the interface. 
The bulk conductance of the heavy fermion can be expressed as
\begin{eqnarray}
G_h={\cal S}\frac{e^2}{h} \frac{p_F^2}{2\pi} \frac{\ell_c}{L} \, ,
\label{gh}
\end{eqnarray}
where $\ell_c$ is $4/3$ of the electron's transport   mean-free path 
and $L$ is the size of the system (in the clean limit, $\ell_c >L$, we set $L/\ell_c=1$ in (\ref{gh})). 
Electron scattering by impurities and phonons are included in the bulk conductances. The point we wish to emphasize is that the 
interface itself causes  a contribution to the total resistance. 

\begin{figure}[htb]
\centerline{\includegraphics[width=7cm]{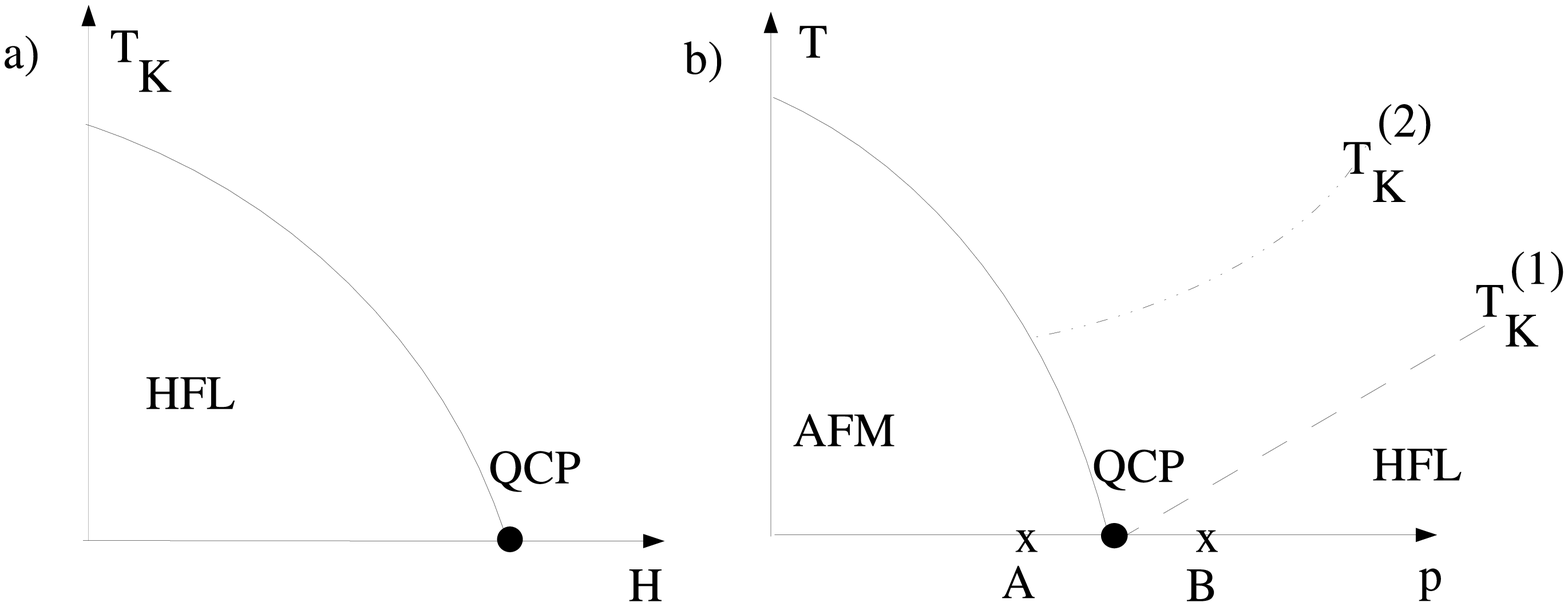}}
\caption{\label{cenarios}  
a) The Kondo temperature decreases under  external magnetic field
and vanishes at a quantum critical point (QCP); b) phase diagram  showing the QCP between 
an antiferromagnetic metal (AFM) and HFL phases, obtained by varying pressure.  
}
\end{figure}

The interface can be created  avoiding
 large lattice mismatches between the two sides, so as to minimize structural scattering. 
Here we suggest a few possibilities: (1) in the scenario of Figure \ref{cenarios}a, 
a large magnetic field can be applied to one side  of the sample and the longitudinal resistivity
can be measured with and without the field; 
(2) a sharp temperature gradient can be applied to a sample of a heavy fermion material so that half of the system is above the
 Kondo temperature and half is below the Kondo temperature.  Heavy fermion materials have resistivity $\rho(T) \propto \log T$ above the 
Kondo temperature $T_K$ and $\rho(T)$ decreases rapidly as temperature is reduced below $T_K$. If a constant temperature difference is
 imposed  across the  sample with thermal conductivity $K$,
 a heat current $j_Q=K\ dT/dx$ flows. From the Wiedemann-Franz law
 we expect $dT/dx \propto T^{-1}\rho(T)$.  If a material with a sharp resistivity peak near the Kondo temperature is chosen, then the 
temperature gradient is high in the region where $T\approx T_K$, producing a thin interface. In this case, the material must
 be chosen so as to minimize effects associated with thermoelectric power;
(3) in the scenario of  Figure \ref{cenarios}b, for systems where the magnetic behavior can be tuned by changing the chemical
 concentration (chemical pressure),  a sample could be
 grown with a large gradient concentration so that half of the system is in the magnetically ordered phase and half in the HFL phase;
(4) alternatively, two different mechanical pressures can be applied on two regions of the sample, 
so that the two regions are on opposite sides of a QCP, as the points A and B in  Figure \ref{cenarios}b. If the QCP 
occurs at vanishing Kondo temperature, $T_K^{(1)}$, the sharp FS volume mismatch at the interface causes the effects
predicted above. 

In systems where the Kondo temperature is finite at the QCP ($T_K^{(2)}$ in  Fig.~\ref{cenarios}(b)), 
we believe the interface scattering between the AFM  and the HFL phases to be negligible small
 (and would anyway disappear as A and B approach the QCP). 
This can be seen as follows. The AFM, in this case, is a spin-density-wave (SDW)
in a system where the local magnetic moments have been Kondo screened by the conduction electrons.
Some regions of the Fermi surface are gapped. 
Now suppose that the electrons flow from the SDW to the paramagnetic phase. 
Only the electrons in the ungapped regions of the Fermi surface produce a current. 
An ungapped electron flows from the SDW into the paramagnetic phase without changing
its Bloch $\bm k$ state. If an interface scattering exists at all, it should be
very small, unlike the one predicted above for the Kondo screening scenario,
which is caused by an appreciable momentum mismatch around the whole Fermi surface.
Therefore, we believe that an experimental setup like the one we propose can
distinguish the two scenarios ($T_K^{(1)}$ and $T_K^{(2)}$  in  Fig.~\ref{cenarios}(b))
for the QCP in heavy fermion systems, because in the SDW scenario the interface
scattering effect should be almost non-existent.

In summary, we have studied the problem of the transport through an interface between a  heavy electron metal and an  ordinary metal. We have argued that the heavy fermion state does not induce a proximity effect on the ordinary metal and that the hybridization gap changes abruptly across the interface. Because of the change of the $f$ electron energy across the interface an atomically thin dipole barrier is formed, a situation quite similar to the dipole layers in semiconducting p-n junctions. We have also shown that the interface produces a contribution to the electrical
 conductance that depends essentially on the number of electrons in the system and that for dense systems the value of the conductance is substantial and can be easily measured experimentally. We have also proposed  ways to create such interfaces. We hope this work will stimulate experimentalists to realize them experimentally. Finally, we would like to stress that our results not only have
 implications for the problem of the nature of the heavy electron ground state, but also can be used to understand quantum criticality and the effects of inhomogeneities close to quantum critical points. 

We thank L. Greene, G. Murthy, and G. Stewart for illuminating conversations.
M.A.N.A.  is grateful to Funda\c{c}\~ao para a Ci\^encia e Tecnologia for a sabbatical grant. A. H. C. N. was supported by the NSF grant DMR-0343790.


\end{document}